\documentclass[sigconf]{acmart}
\renewcommand\footnotetextcopyrightpermission[1]{}
\AtBeginDocument{%
  }

\begin{document}

\title{Integrating AI and Simulation for Teaching Power System Dynamics: An Interactive Framework for Engineering Education}


\author{Osasumwen Cedric Ogiesoba-Eguakun}
\email{oco1411@utulsa.edu}
\authornote{These authors contributed equally as first authors.}
\orcid{0009-0006-8387-058X}
\affiliation{%
  \institution{The University of Tulsa}
  \city{Tulsa}
  \state{Oklahoma}
  \country{USA}
}

\author{Phani Kumar Inkollu}
\authornotemark[1]
\email{phi6517@utulsa.edu}
\affiliation{%
  \institution{The University of Tulsa}
  \city{Tulsa}
  \state{Oklahoma}
  \country{USA}
}

\author{Rupesh Sah}
\authornotemark[1]
\email{rks9560@utulsa.edu}
\orcid{0009-0009-1032-9196}
\affiliation{%
  \institution{The University of Tulsa}
  \city{Tulsa}
  \state{Oklahoma}
  \country{USA}
}

\author{S M Zia Ur Rashid}
\authornotemark[1]
\email{ziaur-rashid@utulsa.edu}
\orcid{0000-0003-3588-5489}
\affiliation{%
  \institution{The University of Tulsa}
  \city{Tulsa}
  \state{Oklahoma}
  \country{USA}
}

\author{Douglas Jussaume}
\email{douglas-jussaume@utulsa.edu}
\affiliation{%
  \institution{The University of Tulsa}
  \city{Tulsa}
  \state{Oklahoma}
  \country{USA}
}

\author{Suman Rath}
\email{suman-rath@utulsa.edu}
\orcid{0000-0002-9012-1919}
\affiliation{%
  \institution{The University of Tulsa}
  \city{Tulsa}
  \state{Oklahoma}
  \country{USA}
}

\renewcommand{\shortauthors}{Ogiesoba-Eguakun et al.}

\begin{abstract}
    Artificial Intelligence (AI), especially cloud platforms and large language models (LLMs), is changing how engineering is taught by making learning more interactive and flexible. However, in electrical engineering and energy systems, students often find power system dynamics difficult to understand because the concepts are abstract, math-heavy, and there are limited opportunities for hands-on practice. This paper presents an AI-based interactive learning framework that combines simulation with intelligent feedback to improve understanding and student engagement. The framework has three connected parts: an AI layer that provides explanations and guidance, a simulation layer that models system behavior, and a user layer that allows students to interact with the system in real time. These parts work together in a continuous loop where students explore how the system behaves, change parameters, and receive feedback based on the results. The paper also provides a step-by-step process to help educators design and apply AI-supported learning environments, including breaking down concepts, using simulations, and assessing performance. This method helps students learn through practice and better understand how ideas from class apply to real power systems. It also provides a practical way to improve electrical engineering education and helps students get ready to use AI tools carefully and responsibly in engineering.
\end{abstract}

\ccsdesc[500]{Applied computing~Education}
\ccsdesc[300]{Computing methodologies~Machine learning}
\ccsdesc[300]{Applied computing~Interactive learning environments}

\keywords{Artificial Intelligence, Engineering Education, Microgrids, Interactive Learning, Simulation-Based Learning, Large Language Models}

\maketitle

\section{Introduction}
Electrical engineering education, especially in power and energy systems, deals with complex and fast-changing processes like system stability, control, and cyber-physical interactions \cite{van2017cyber}. These topics are often hard to understand because they are abstract and involve a lot of mathematics. This makes it hard for students to understand what is happening in the system, especially in traditional lecture-based classes \cite{radianti2020systematic}.
One major challenge is helping students connect theory with what actually happens in real power systems. Many students find it hard to understand how system variables change during disturbances, how control actions affect stability, and how different parts of the system work together over time. This problem becomes worse because access to lab equipment is limited, and building or maintaining such setups is expensive, which reduces opportunities for hands-on learning. To solve this, many courses use simulation tools and lab exercises. While these help, they often lack interaction, flexibility, and real-time support. Students often follow set steps with little chance to explore or learn by trying things on their own, which can lead to a weak understanding \cite{radianti2020systematic}. Recent advances in AI, especially large language models and cloud tools, are giving new ways to make engineering education better and more interactive \cite{zawacki2019systematic,holmes2019artificial}. However, these tools are still not well organized or properly combined with simulation-based learning \cite{holmes2019artificial}. This paper presents an AI-based interactive learning framework that combines simulation with intelligent feedback to support learning through practice in power system education. The framework connects AI explanations, system simulation, and user interaction in a continuous learning loop \cite{tao2018digital}, helping students better understand concepts while also building practical skills.

Table~\ref{tab:problem_solution} presents the key challenges in traditional power system education and the corresponding solutions provided by the proposed framework.

\begin{table}[ht]
\caption{Mapping of key challenges to the proposed AI-driven solution.}
\label{tab:problem_solution}
\centering
\begin{tabular}{p{3.0cm} p{4.0cm}}
\toprule
\textbf{Challenge} & \textbf{Proposed Solution} \\
\midrule
Abstract concepts & Use simulation and visualization to improve understanding \\

Limited hands-on learning & Provide interactive simulation-based environment \\

Passive learning & Enable active exploration through user interaction \\

Weak theory-practice link & Connect concepts with real-time system behavior \\

Lack of guidance & Integrate AI for feedback and explanation \\
\bottomrule
\end{tabular}
\end{table}

\section{AI-Driven Interactive Learning Framework}

\subsection{Framework Overview and Architecture}
\begin{figure}[t]
\centering
\includegraphics[width=\columnwidth]{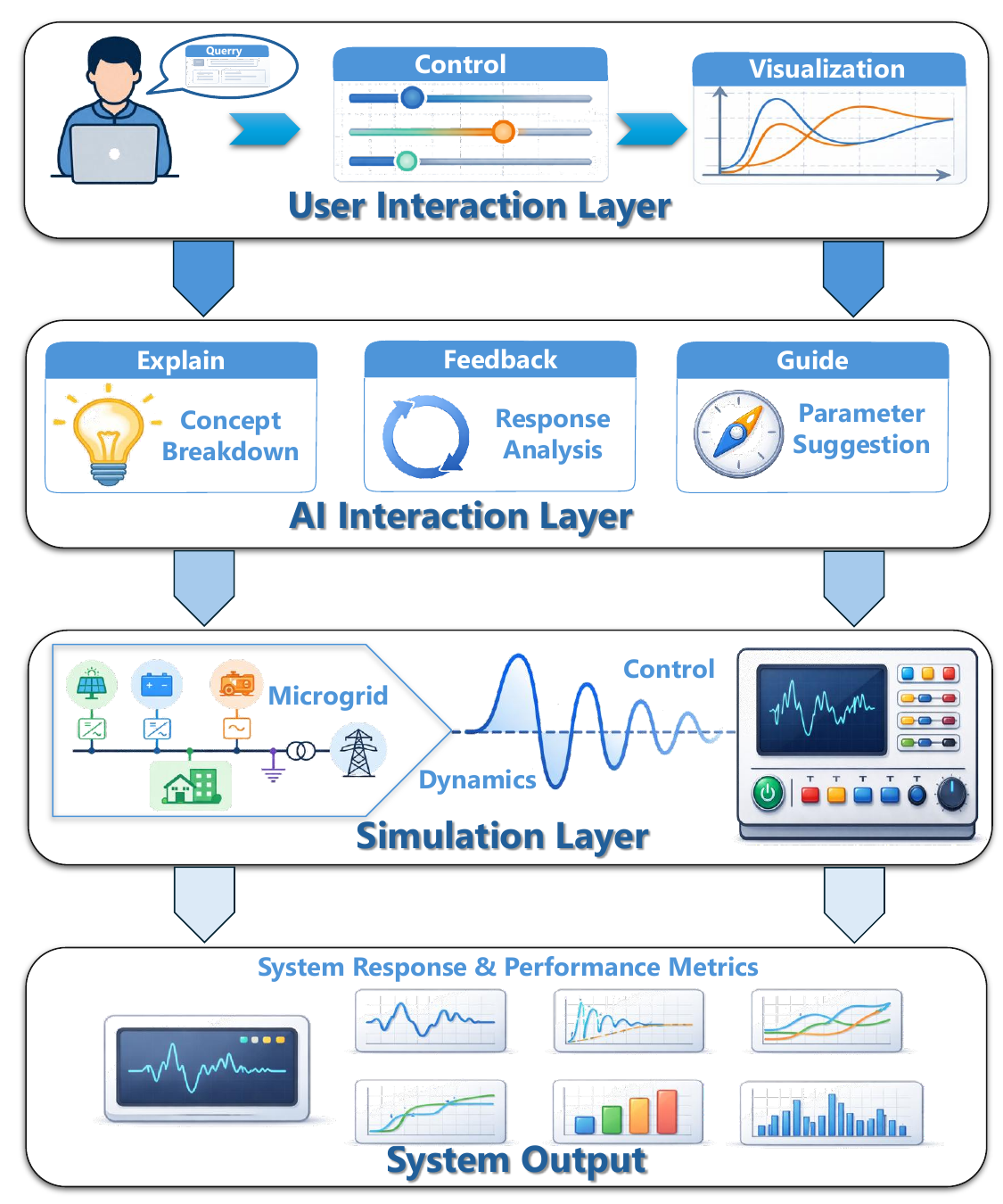}
\caption{Proposed AI-driven interactive learning framework for power system dynamics. The system connects user interaction, AI guidance, and simulation models in a continuous loop to support simple, hands-on learning.}
\label{fig:framework}
\end{figure}

The proposed framework allows students to learn power system dynamics interactively by combining AI with simulation tools. It is designed as a closed-loop system where user interaction, simulation, and AI feedback work together. This setup allows students to explore how the system behaves step by step while receiving real-time guidance. As they test different ideas and see the results, their understanding improves through practice. The framework has three connected parts: the \textit{AI Interaction Layer}, the \textit{Simulation Layer}, and the \textit{User Interaction Layer}. The full structure of the framework is shown in Fig.~\ref{fig:framework}.
The User Interaction Layer is the interface where students ask questions, adjust system parameters, and observe the results, supporting hands-on learning. The AI Interaction Layer uses large language models to help explain concepts, guide students, and give adaptive feedback. It simplifies complex concepts, explains system behavior, and recommends parameter adjustments to improve understanding.  The Simulation Layer models how power systems behave over time, including microgrid operation, changes in frequency and voltage, and control actions during disturbances. It allows real-time changes to parameters and visualizes how the system behaves \cite{tao2018digital,van2017cyber}. 

\subsection{Learning Workflow and Implementation}

The framework operates through a continuous interaction workflow that transforms passive learning into an active process. A student begins by initiating a learning task or concept query, after which the AI defines the simulation context and provides an initial explanation. The simulation model is then run, and the student interacts with it by changing inputs or parameters. As the results update to show how the system behaves, the AI explains what is happening and provides feedback. Through this repeated process, students explore the system and build a better understanding through hands-on interaction. The implementation follows a structured methodology. At first, key topics like frequency stability and power flow are chosen for interactive learning. Then, simulation models are used to show how the system behaves under different conditions, and AI tools are connected to explain the results and provide feedback. This is followed by the design of user interfaces that support real-time interaction and visualization. Finally, how well students understand the concepts and apply them in practice is used to assess their learning.

\subsection{Case study: Microgrid Frequency Control}

Microgrid frequency control is used as an example to show how the framework works. Students interact with a simulated microgrid environment in which disturbances such as load variations are introduced. By adjusting controller parameters, such as proportional and integral gains, students can see how the system responds. The AI system explains how these changes affect system stability, damping, and steady-state performance. Students gradually develop both a practical and a deeper understanding of frequency control in power systems through repeated interaction.

\section{Conclusion and Future Work}
This paper introduces a framework that uses AI to improve electrical engineering education through interactive and simulation-based learning. By integrating AI tools into teaching practices, educators can improve student engagement, comprehension, and practical skills. Future work will focus on building full virtual labs, adding digital twin systems, and testing how well AI-supported learning works in larger classroom settings. In addition, the research will explore the ethical use of AI in education and ways to ensure it is used responsibly.


\bibliographystyle{ACM-Reference-Format}
\bibliography{sample-base}

@String{Springer = "Springer-Verlag" }

@article{zawacki2019systematic,
  title={Systematic review of research on artificial intelligence applications in higher education--where are the educators?},
  author={Zawacki-Richter, Olaf and Mar{\'\i}n, Victoria I and Bond, Melissa and Gouverneur, Franziska},
  journal={International journal of educational technology in higher education},
  volume={16},
  number={1},
  pages={39},
  year={2019},
  publisher={Springer}
}

@book{holmes2019artificial,
  title={Artificial intelligence in education promises and implications for teaching and learning},
  author={Holmes, Wayne and Bialik, Maya and Fadel, Charles},
  year={2019},
  publisher={Center for Curriculum Redesign}
}

@article{tao2018digital,
  title={Digital twin in industry: State-of-the-art},
  author={Tao, Fei and Zhang, He and Liu, Ang and Nee, Andrew YC},
  journal={IEEE Transactions on industrial informatics},
  volume={15},
  number={4},
  pages={2405--2415},
  year={2018},
  publisher={IEEE}
}

@article{radianti2020systematic,
  title={A systematic review of immersive virtual reality applications for higher education: Design elements, lessons learned, and research agenda},
  author={Radianti, Jaziar and Majchrzak, Tim A and Fromm, Jennifer and Wohlgenannt, Isabell},
  journal={Computers \& education},
  volume={147},
  pages={103778},
  year={2020},
  publisher={Elsevier}
}

@inproceedings{van2017cyber,
  title={Cyber-physical energy systems modeling, test specification, and co-simulation based testing},
  author={van der Meer, Arjen A and Palensky, Peter and Heussen, Kai and Bondy, DE Morales and Gehrke, Oliver and Steinbrinki, Cornelius and Blanki, Marita and Lehnhoff, Sebastian and Widl, Edmund and Moyo, Cyndi and others},
  booktitle={2017 Workshop on Modeling and Simulation of Cyber-Physical Energy Systems (MSCPES)},
  pages={1--9},
  year={2017},
  organization={IEEE}
}

\appendix

\end{document}